\documentclass[preprintnumbers,amsmath,nofootinbib, amssymb,latexsym,floatfix,superscriptaddress,onecolomn,prd]{revtex4}

\usepackage{graphicx}
\usepackage{dcolumn}
\usepackage{bm}

\usepackage{color,xcolor}

\makeatletter    

\begin{document}

\title{Schwinger effect impacting primordial magnetogenesis}

 \author{Cl\'{e}ment~Stahl}\email{clement.stahl@pucv.cl}
\affiliation{Instituto de Fisica, Pontificia Universidad Catolica de Valparaiso, Casilla 4950, Valparaiso, Chile} 

\begin{abstract}
We explore the enhancement of an electromagnetic field in an inflationary background with an anti-conductive plasma of scalar particles. The scalar particles are created by Schwinger effect in curved spacetime and backreact to the electromagnetic field. The possibility of a negative conductivity was recently put forward in the context of the renormalization of the Schwinger induced current in de Sitter spacetime. While a negative conductivity enhances the produced magnetic field, we find that it is too weak to seed the observed intergalactic magnetic field today. This results on pair creation in inflationary scenario is however important for primordial scenarios of magnetogenesis as the presence of a conductivity alters the spectral index of the magnetic field. This also shows on a specific example that backreaction can increase the electromagnetic field and not only suppress it.
\end{abstract}

\date{\today}

\maketitle

\section{Introduction}
\par
The generation of large scales magnetic fields in our present universe is an open problem of modern cosmology. Astrophysical magnetic fields of $\mu$G amplitude have been observed in galaxies, clusters and high redshift ($z<4$) objects. The observed correlation is of the order of the object size making it therefore hard to explain. To generate galactic fields of $\mu$G amplitude, by flux conservation during
the formation of the galaxy, seed fields of about $10^{-9}$ G are required. Besides, coherent cosmological magnetic fields have also been observed on Mpc scales: the observation of blazars with gamma-ray telescopes gave a lower bound of $B_0 > 10^{-15}$ G for the present magnetic field \cite{Neronov:1900zz}. Two broad classes of model to generate those seeds exist. The first: an astrophysical origin is a natural possibility: after recombination, plasma effects, exploiting the difference in mobility between electrons and ions, generate currents and thus magnetic field through Maxwell equations. Those mechanisms are dubbed under the generic term of batteries further amplified by flux freezing and dynamo action times mechanism \cite{Brandenburg:2004jv,Durrer:2013pga}, a classical example is the Biermann battery. The second: a primordial origin: \cite{Grasso:2000wj,Kandus:2010nw,Durrer:2013pga,Subramanian:2015lua} could explain both the observation of magnetic fields in all the structures and the presence of cosmological magnetic fields in the intergalactic medium. A combination of both scenarios (an astrophysical and a primordial origin of the magnetic fields) is also a possibility.
\par
In the primordial universe, either local processes such as phase transitions \cite{Vachaspati:1991nm,Sigl:1996dm}, magnetohydrodynamics turbulence, the presence of charge and current density together with vorticity, or non local processes during inflation could generate seed magnetic fields. Via local processes, usually, the correlation length is found to be too small and the generation of a blue power spectrum leads to too small seeds \cite{Durrer:2003ja}. On the other hand non-local processes happening during inflation generate magnetic fields at all scales \cite{Turner:1987bw,Ratra:1991bn,Gasperini:1995dh,Giovannini:2000dj,Davis:2000zp,Bamba:2003av,Anber:2006xt,Martin:2007ue,Demozzi:2009fu,Durrer:2010mq,Ferreira:2013sqa,Ferreira:2014hma}, however as electromagnetism is conformally invariant, the generated $B$ field decrease as $a^{-2}$ rendering difficult to have strong fields. It is possible to break the conformal invariance of the Maxwell field to obtain the necessary observed field strength.
\par
The conformal invariance was broken by introducing a tachyonic mass to the photon which may arise from a non-minimal coupling to gravity \cite{Turner:1987bw}, however such theories usually are not physically relevant because of the appearance of ghosts \cite{Demozzi:2009fu,Himmetoglu:2009qi}. The axion model \cite{Garretson:1992vt} is also a possible alternative where one couples a rolling pseudo-scalar to the photon field. It is characterized by a very blue spectrum ($P \propto k^2$) that usually rules those models out. However, those models have gained some interest in the past years. Indeed, the produced fields are highly helical and could undergo, during the matter and radiation era, a process of inverse cascade that transfers the power from small scale to large scales \cite{Anber:2006xt,Banerjee:2004df,Caprini:2014mja}.
One other popular way to break the conformal invariance of the Maxwell field is to introduce a non-minimal kinetic coupling of the inflaton to the EM field dubbed the Ratra model \cite{Ratra:1991bn} or IFF models. This idea was revisited in \cite{Martin:2007ue} in the context of string-inspired inflation models and also considered in the context of dilaton electromagnetism \cite{Bamba:2003av} and DBI (Dirac-Born-Infeld) inflation \cite{Bamba:2008my}. Many of these models are plagued by two main problems: the backreaction problem and the strong coupling problem. The first one arises, when one obtains too large electromagnetic (EM) fields that could impact the global inflationary dynamics and spoil the inflation \cite{Demozzi:2009fu,Kanno:2009ei}. A high electric field during inflation would indeed modify also the amplification of scalar perturbation during inflation and would have a direct impact on the Cosmic Microwave Background \cite{Barnaby:2012tk} thus constraining magnetogenesis scenarios \cite{Bonvin:2013tba,Fujita:2014sna,Ferreira:2014hma}. The second one arises on a technical level when the conformal invariance is broken leading to a large effective EM coupling rendering the perturbative calculations for the EM field untrustable \cite{Gasperini:1995dh,Demozzi:2009fu}. 
An interesting escape to the strong coupling is to drop gauge invariance \cite{Bonvin:2011dt} however it is difficult to write a model where the gauge invariance is restored once broken. Mechanism that break gauge invariance generally also give rise to ghost-like instabilities \cite{Himmetoglu:2008zp,Himmetoglu:2009qi}. Another way to escape this problem is to lower the energy scale of the inflation as done in \cite{Ferreira:2013sqa} where a Tev scale inflation would generate the observed present magnetic field. 
\par  A corollary of those problems is the generation of Schwinger pairs screening the establishment of a large magnetic field \cite{Kobayashi2014,Sharma:2017eps}. Schwinger mechanism  \cite{Sauter:1931zz,Heisenberg1936,Schwinger1951,Gelis:2015kya} is the spontaneous creation of pairs of particle/anti-particles in the presence of a strong electromagnetic field. Its generalization in de Sitter space (dS) has attracted much attention in the past years \cite{Garriga:1993fh,Kim:2008xv,Froeb2014,Kobayashi2014,Stahl:2015gaa,Stahl:2016geq,Hayashinaka:2016qqn,Hayashinaka:2016dnt,Yokoyama:2015wws,Geng:2017zad,Sharma:2017ivh,Sharma:2017eps,Bavarsad:2017oyv,Giovannini:2018qbq,Hayashinaka:2018amz,Ferreiro:2018qdi,Banyeres:2018aax,Chua:2018dqh}. In those setups both the gravitational field and a strong EM field are generating pairs out of the quantum vacuum. Those pairs then move along the electric field generating currents and thus a positive conductivity which prevents the amplification of the EM field. In the strong electric field regime, it was found that this Schwinger current constraints magnetogenesis scenarios and an upper bound on the current magnetic field was derived \cite{Kobayashi2014}. Recently, it was found that the conductivity of the spacetime could be negative in the weak electric regime \cite{Hayashinaka:2018amz}, leading possibly to an enhancement of the electromagnetic field by an anti-screening mechanism. While most of the studies \cite{Garriga:1993fh,Kim:2008xv,Froeb2014,Kobayashi2014,Stahl:2015gaa,Hayashinaka:2016qqn,Sharma:2017eps,Ferreiro:2018qdi,Chua:2018dqh} assume that the Schwinger pairs are created by a constant electric field, it was shown recently \cite{Giovannini:2018qbq} that such a configuration is violating the second law of thermodynamics. It is therefore desirable to investigate more realistic field configurations. The purpose of this article is to go in this direction by considering the backreaction of the Schwinger pairs to the electromagnetic field. We will focus on a setup where the conductivity of the spacetime is negative (see equation \eqref{eq:current}). Such as possibility leads to an instability and effectively, the conformal invariance is broken by the backreaction of the quantum effects. In turn, we will see that as soon as EM field is included in an inflationary setup, the Schwinger effect needs to be taken into account. Similar steps are made in the same direction in \cite{Kitamoto:2018htg,Sobol:2018djj}. Motivations to consider a negative conductivity may comes from adopting the ``maximal subtraction'' renormalization scheme of \cite{Hayashinaka:2018amz} or from the nonlinear corrections to the Maxwell action and the logarithmic running of the coupling constant in \cite{Banyeres:2018aax}.
\par The plan of this article is as follows: in section \ref{sec:SchwingerindS}, we review the latest results on Schwinger effect in de Sitter space following the presentation of \cite{Yokoyama:2015wws}. In section \ref{sec:inflmagn}, we present the basic equations for our setup and calculate the electromagnetic power spectra at the end of inflation. Eventually in section \ref{sec:discussion}, we give some concluding remarks and discuss some possible extensions of this work.
\section{Schwinger effect in de Sitter space}
\label{sec:SchwingerindS}
We consider the production of charged scalar particles $\varphi$ of mass $m$ in $1+3$ D under the influence of a constant electric field in dS. Only in this section, we consider that the scale factor $a$ of the de Sitter metric reads:
\begin{align}
\label{eq:dS}
a(\eta)= - \frac{1}{1-H\eta},
\end{align}  
where $\eta$ is the conformal time running from $-\infty$ to $H^{-1}$ and $H$ is the Hubble constant. The gauge field $A_{\mu}$ ($\mu=0,1,2,3$) is considered pointing in the z-direction with a constant amplitude $E^2$. We will discuss in more detail in section \ref{sec:inflmagn} the relevance of modeling the electric field as a constant one. In dS, (\ref{eq:dS}), it reads:
\begin{align}
A_{\mu}=- \frac{E}{H}\left(a(\eta)-1\right) \delta_{\mu}^{z}.
\end{align}
With such as choice, the Lorentz and Coulomb gauge conditions are satisfied: $A_0=\partial_i A_i = 0$.
In order to quantize, the rescaled complex scalar field $\chi=a  \varphi $ is expanded on modes function:
\begin{align}
\chi(\eta,\textbf{x})=\int \frac{d^3 \textbf{k}}{(2 \pi)^3} e^{i \textbf{x}.\textbf{k}}\left(\chi_{\textbf{k}}(\eta)b_{\textbf{k}}+\chi_{\textbf{k}}^{*}(\eta)d_{-\textbf{k}}^{\dagger} \right)
\end{align} 
The commutation relation for the annihilation and creation operators are:
\begin{equation}
\left[b_{\textbf{k}},b_{\textbf{k}'}^{\dagger} \right]=\left[d_{\textbf{k}},d_{\textbf{k}'}^{\dagger} \right]=(2\pi)^3 \delta^{(3)}(\textbf{k}-\textbf{k}'),
\end{equation}
the other commutators being 0. The Klein-Gordon equation for each modes reads:
\begin{equation}
\label{eq:KG}
\left[\partial_{\eta}^2 + 
\omega_{\textbf{k}}^2(\eta) \right]\chi_{\textbf{k}}=0,
\end{equation}
where the effective pulsation is: $\omega^2=\textbf{p}^2+\frac{\beta}{1-H \eta}+\frac{\gamma}{(1-H \eta)^2}$, the shifted momentum is: $\textbf{p}=\textbf{k}+(0,0,\frac{eE}{H})$, $\alpha=m^2+\left(\frac{eE}{H}\right)^2$ and $\beta=-2\frac{eE}{H}p_z$.
Changing the time variable to $z=-2ip\left(\frac{1}{H}-\eta \right)$, two independent solutions to (\ref{eq:KG}) are the Whittaker functions $\mathcal{W}_{\kappa, \mu } $ and $\mathcal{M}_{\kappa, \mu }$ which are hypergeometric functions. Further defining $\mu^2 = 9/4 - M^2 -L^2$, $\kappa=-iL \frac{p_z}{p}$ and the dimensionless electric field and mass: $M=\frac{m}{H}$, $L=\frac{eE}{H^2}$, it is then possible to define the in-modes with only positive frequency in the asymptotic past ($\eta \rightarrow -\infty$):
\begin{equation}
\label{eq:inmodes}
\chi_{\textbf{k}}^{\text{in}}(\eta)=\frac{e^{i \kappa \pi/2}}{\sqrt{2p}}\mathcal{W}_{\kappa, \mu }(\eta).
\end{equation}
For $\mu$ imaginary, the out-modes, in the asymptotic future ($\eta \rightarrow H^{-1}$) are:
\begin{equation}
\label{eq:outmodes}
\chi_{\textbf{k}}^{\text{out}}(\eta)=\frac{e^{i |\mu| \pi/2}}{\sqrt{4|\mu|p}}\mathcal{M}_{\kappa, \mu }(\eta)
\end{equation}
Performing the standard Bogoliubov calculation requires the connexion formula linking the two Whittaker functions: $\mathcal{W}_{\kappa, \mu } $ and $\mathcal{M}_{\kappa, \mu }$:
\begin{equation}
\label{eq:connexion}
\mathcal{W}_{\kappa, \mu } = \frac{\Gamma(-2 \mu)}{\Gamma(1/2-\mu- \kappa)}\mathcal{M}_{\kappa, \mu } + \frac{\Gamma(2 \mu)}{\Gamma(1/2+\mu- \kappa)}\mathcal{M}_{\kappa, -\mu }.
\end{equation}
With (\ref{eq:inmodes}), (\ref{eq:outmodes}) and (\ref{eq:connexion}), the Bogoliubov coefficient can be obtained, they read:
\begin{align}
& \alpha_{\textbf{k}}=\frac{\sqrt{2 |\mu|} \Gamma(-2\mu)}{\Gamma(1/2-\mu -\kappa)}e^{i (\kappa + |\mu|) \pi/2} \\
& \beta_{\textbf{k}}=-i\frac{\sqrt{2 |\mu|} \Gamma(2\mu)}{\Gamma(1/2+\mu -\kappa)}e^{i (\kappa - |\mu|) \pi/2}.
\end{align}
The number of particles in the asymptotic future is given by:
\begin{equation}
\label{eq:paircrea}
N_{\textbf{k}}=|\beta_{\textbf{k}}|^2=\frac{e^{2 \pi L p_z/p}+e^{-2 \pi|\mu|}}{e^{2 \pi |\mu|}-e^{-2\pi |\mu|}}.
\end{equation}
In term of the effective action: $N_{\textbf{k}}=\exp(-S)$, two limits can be discussed: in the strong electric field regime ($L \gg 1$), one recovers the Schwinger formula : the flat spacetime result which reads in therm of the action: $S=\pi m^2/(eE)$. In the weak electric field regime ($L \ll M$), one finds: $S \sim 2 \pi (M-|L|)$: the gravitational pair creation is dominating over its electric counter part. The first term in the effective action is a Boltzmann factor for non-relativistic particles at Gibbon-Hawking temperature, while the second term is the corrections to the first term arising from the small electric field. Observe that unlike the flat spacetime result, in (\ref{eq:paircrea}), it is possible to create pairs with momentum opposite to the electric field sense: with $p_z<0$ however in the flat spacetime limit those modes gets exponentially suppressed.
\par Using a semiclassical approximation, it is possible to calculate the total number of pairs in inflationary setups \cite{Kobayashi2014}:
\begin{equation}
n=\frac{|\mu|^3 \sinh(2 \pi L)}{3 L\sinh(2 |\mu| \pi)} \frac{H^3}{(2 \pi)^3}
\end{equation}  
This result is constant even though the background expands exponentially. It comes from the fact that the electric field has been taken to be constant also. At each Hubble time, pairs get created via Schwinger effect. In this way the population of $\varphi$ particles is always dominated by those created within a Hubble time. We will discuss in section \ref{sec:inflmagn}, the range of application of this approximation for the purpose of magnetogenesis.
\par This semiclassical calculation (only valid for $|\mu| \gg 1$) motivates the exploration of a larger parameter space. The created particles will move along the electric field and create a current and therefore a conductivity. The current can be derived without the assumption of an asymptotic future for the particle and therefore is a more general quantity. It is defined as:
\begin{equation}
J_{\mu}=ie \left[(D_{\mu} \varphi)^{\dagger}\varphi-\varphi^{\dagger}D_{\mu} \varphi \right],
\end{equation}
where $D_{\mu}=\partial_{\mu}+ieA_{\mu}$. This current can be shown to be conserved and the only non-vanishing component of it is the z-component. It is possible to calculate it in term of the solutions to the Klein-Gordon equation (\ref{eq:KG}):
\begin{equation}
<J_{i}>=\frac{2e}{a^2} \int \frac{d^3 \textbf{k}}{(2 \pi)^3} (k_i+eA_i)|\chi_{\textbf{k}}^{\text{in}}|^2.
\end{equation}
It is a divergent quantity which needs to be renormalized. The choice of the renormalization scheme is not a trivial one and different possibilities have been explored in the past years. The original work \cite{Kobayashi2014} used the adiabatic subtraction up to order 2 while in \cite{Hayashinaka:2016dnt}, it has been shown that the point-splitting technique was giving the same final result. We will mention also the Pauli-Villars scheme which is equivalent to adiabatic subtraction in the equivalent problem in $1+1$ D \cite{Froeb2014}. However the final result was exhibiting a bizarre behavior dubbed infrared-hyperconductivity \cite{Froeb2014}: for a decreasing cause: the electric field, the consequence: the induced current was increasing. Besides, it was also noted in \cite{Hayashinaka:2016qqn} that in the massive limit $M \gg 1$, the bosonic and the fermionic current disagrees despite the expected agreement on the semiclassical level \cite{Stahl:2015cra}. Motivated by these facts, a new renormalization scheme was proposed in \cite{Hayashinaka:2018amz} dubbed maximal subtraction. The interested reader may look at equation (14) of \cite{Hayashinaka:2018amz} to see the difference between adiabatic (minimal) subtraction and maximal subtraction. One of the consequence of maximal subtraction is the appearance of a negative conductivity when an electric field is present. This negative conductivity was already reported in \cite{Hayashinaka:2016qqn} for the case of fermionic particles. In the case of bosonic particles, defining the dimensionless renormalized current: $\mathcal{J}=\frac{J^{z}_{\text{ren}}}{ea^3 H^3}$, it has been found \cite{Hayashinaka:2018amz} that in the massive limit, the current has the following expression:
\begin{equation}
\label{eq:negcurr}
\mathcal{J}=-\frac{8}{9 \pi} M^3 \exp(-2\pi M) L.
\end{equation}  
The issue of renormalization schemes in curved spacetime is thorny: we mention the very recent works \cite{Ferreiro:2018qzr,Ferreiro:2018qdi} where the adiabatic subtraction method was adopted and the conformal anomaly was generalized to the case of a constant electric field in curved de Sitter space. In \cite{Ferreiro:2018qdi}, the limiting cases of pure electric field and pure gravitational fields for the conformal anomaly are recovered in the relevant limit however to do so they assumed that the vector potential was taken to be of adiabatic order 1 instead of 0 as in the previous works \cite{Kobayashi2014,Stahl:2015gaa}. In \cite{Ferreiro:2018qzr}, in the same setup, it was checked whether in that case the total energy-momentum is conserved. In conclusion, the choice of a renormalization scheme for setups involving test particles, an electromagnetic field and a gravitational field is still work in progress and motivate us to investigate physical applications such as the possibility of a negative conductivity for primordial magnetogenesis scenarios.
\par 
Around equation (4.9) of \cite{Banyeres:2018aax}, a new interpretation of equation \eqref{eq:negcurr} is proposed. The infrared behavior of the current is understood as non-linearities of the kinetic term in the electromagnetic field. The Euler-Heisenberg Lagrangian in curved space-time generalizes the Maxwell one and the charge $e$ gets a logarithmic running. Such terms generate also a negative current if the logarithmic running is large enough.
\section{Inflationary magnetogeneis}
\label{sec:inflmagn}
Our model consists in a scalar field minimally coupled to a gauge field.
\begin{equation}
S= \int \sqrt{-g} d^4 x \left[-\frac{1}{4 \pi} F_{\mu \nu} F^{\mu \nu}+g^{\mu \nu}(\partial_{\mu}-ieA_{\mu}) \varphi^*(\partial_{\nu}+ieA_{\nu}) \varphi-m^2 \varphi^2 \right],
\end{equation}
where the field strength is $F_{\mu \nu} = \partial_{\mu} A_{\nu}-\partial_{\nu} A_{\mu}$ and $A_{\mu}$ is the 4-potential of the EM field. We use the following conventions: the greek indices $\mu, \nu$ are spacetime indices and run from 0 to 3 whereas the latin indices $i,j$ are space indices and run from 1 to 3. The signature of the metric is $(+ - - -)$. $g$ is the de Sitter metric which reads in conformal coordinates:
\begin{equation}
\label{eq:metric}
ds^2=a(\eta)^2 (\text{d}\eta^2-\text{d}x^2-\text{d}y^2-\text{d}z^2).
\end{equation}
The conformal time is now parametrized by the Hubble factor in the following way: \begin{equation}\eta= -\frac{1}{aH},\hspace{1cm} a^2 H \equiv \frac{\text{da}}{\text{d} \eta }, \hspace{1cm} (-\infty<\eta<0).
\end{equation}
Using the results of section \ref{sec:SchwingerindS} for Schwinger effect dS, at each Hubble time, pairs are created. In this picture, the typical time of pair creation $t_{\text{pair}}=\frac{1}{H}$ has to be much smaller that the typical time of evolution of the electromagnetic field $t_{E} = \frac{1}{k}$. Therefore, focusing on the large scales it is possible to consider the electric field as constant. We comment here on the results of \cite{Giovannini:2018qbq} that states the to sustain a constant electromagnetic field in dS requires to have Maxwell currents that violates the second law of thermodynamics. The approximation of a constant electromagnetic is a toy model which holds for large scales. In the more realistic setup of this article, we expect the second law of thermodynamics, which is nothing but microcausality: a local process, to hold. Besides, as we investigate the possibility of a negative conductivity as in equation (\ref{eq:negcurr}) which appears in the weak electric field limit: $L\ll 1$, this avoids, in passing, the backreaction problem. We assume that a Maxwell current of the form
\begin{equation}
J_i= H \sigma E_i
\label{eq:current}
\end{equation}
is present in our setup. Here $\sigma$ is a dimensionless negative constant with respect to time. It depends on the details of the pair creation process. The conformal invariance prevents perturbations of a gauge-field to develop, however here, the Schwinger mechanism breaks it with quantum fluctuations inducing a current and leading to the possibility of parametric amplification of the gauge field. The case of a strong electric field leads to a positive conductivity which screens the amplification of the EM field \cite{Kobayashi2014,Stahl:2016geq}.
\subsection{Electric and magnetic fields}
We consider $u^{\mu}=\left(\frac{1}{a(\eta)}, \vec{0}\right)$ the 4-velocity of a comoving observer and $\epsilon_{\mu \nu \alpha \beta} \equiv \eta_{\mu \nu \alpha \beta} \sqrt{-g}$ is the curved spacetime Levi-Civita tensor, $\eta_{\mu \nu \alpha \beta}$ is the corresponding one in flat spacetime. Our convention is $\eta_{0123} =1$. In this framework, the components of the electromagnetic field read:
\begin{align}
E_{\mu}(\eta) & = u^{\nu}F_{\nu\mu}, \label{eq:EfonctionofA} \\
B_{\mu}(\eta) & = \frac{1}{2} \epsilon_{\nu\mu\rho\sigma}u^{\nu}F^{\rho\sigma}
\end{align}
We work in the Coulomb gauge:
\begin{equation}
\label{eq:coulomb}
 A_{i,i}= A_{\eta}=0.
\end{equation}
Using (\ref{eq:coulomb}) together with the metric (\ref{eq:metric}), one can show that the time component for the electric and magnetic field vanishes while the space components read:
\begin{align}
& E_i = -\frac{1}{a} A'_i, \\
& B_i = \frac{1}{a^4} \epsilon_{ijk0} u^0 \partial_j A_k,
\end{align}
while the magnitudes of the fields are
\begin{align}
& E^2 \equiv E_{\mu} E^{\mu}= \frac{1}{a^4} A_i' A_i', \\
& B^2 \equiv B_{\mu} B^{\mu} = \frac{1}{a^4} \left(\partial_i A_j \partial_i A_j - \partial_j A_i \partial_j A_i \right).
\end{align}
\subsection{Backreaction} In the presence of a source, the inhomogeneous Maxwell equation reads:
\begin{equation}
\label{eq:maxwellfree}
\nabla^{\nu}(F_{\mu \nu})= a^2 J_{\mu}.
\end{equation}
Using the Coulomb gauge (\ref{eq:coulomb}), equation (\ref{eq:maxwellfree}) gives:
\begin{equation}
\label{eq:eomfree}
A_i''-\partial_j \partial_j A_i =\frac{\sigma}{\eta} A'_i,
\end{equation}
where a prime denotes a derivative with respect to the conformal time $\eta$.
Observe that the addition of $\sigma$ gives the same type of equations that the ones obtained in IFF scenarios. Therefore our model can be seen as an effective way to break the conformal coupling or as a way to justify it with a physical process: Schwinger effect in curved spacetime. We stress again that this analogy holds for $L \ll 1$ while the $L \gg 1$ case was treated in \cite{Kobayashi2014}.
\par From (\ref{eq:eomfree}) describing the gauge potential $A_i$, we rotate to Fourier space and consider $A(k,\eta)$. We finally change the gauge variable to the canonical variable for quantization: $\mathcal{A}(k,\eta) \equiv a^{\sigma/2} A(k,\eta)$. The equation of motion reads:
\begin{equation}
\label{eq:eom}
\mathcal{A}''(k,\eta)+\left[k^2-H^2 a^2 \left(\frac{\sigma}{2}\right) \left( 1-\frac{\sigma}{2}\right) \right] \mathcal{A}(k,\eta)=0
\end{equation}
which is the equation of an harmonic oscillator with a time dependent frequency. The detailed calculation for the quantization procedure can found in \cite{Martin:2007ue} remembering the fact that the Schwinger effect is equivalent in the equation of motion to a IFF scenario.
\subsection{Power spectrum} A solution for (\ref{eq:eom}) exists in term of Hankel function of the first kind: $H^{(1)}_{\nu}$:
\begin{equation}
\label{eq:sol}
\mathcal{A}(k,\eta)= \frac{\sqrt{\pi}}{2} \sqrt{-\eta} H^{(1)}_{\frac{\sigma+1}{2}}(-k \eta),
\end{equation}
where the mode function $\mathcal{A}(k,\eta)$ has been normalized such that $\lim_{k \eta\to-\infty}\mathcal{A}(k,\eta) =\frac{1}{\sqrt{2k}} \exp(-i k \eta)$, up to an arbitrary phase.
Using the fact widely used in inflationary cosmology that at late time, the vacuum expectation of the fields generated during inflation becomes classical and can be interpreted as stochastic power spectra, we are aiming at computing the magnetic power spectrum defined as:
\begin{align}
& P_E \equiv \frac{k^3 |\mathcal{A'}|^2}{\pi^2 a^4} \\
& P_B \equiv \frac{k^5 |\mathcal{A}|^2}{\pi^2 a^4}
\end{align}
We stress that any electric field produced during inflation will decay during the reheating phase where the universe is filled with relativistic standard model particles: a relativistic highly conductive plasma. In that case, MHD approximation holds and the electric field will dissipate and not be present anymore. Taking (\ref{eq:sol}) in the asymptotic future ($|k\eta| \ll 1$), we find the electric and magnetic power spectra at the end of inflation to be proportional to:
\begin{align}
& P_E \propto H^4 \left(\frac{k}{aH} \right)^{\sigma+4} \\
& P_B \propto H^4 \left(\frac{k}{aH} \right)^{5-|\sigma +1|}
\end{align}
The case $\sigma=-6$ gives a scale invariant power spectrum: $P_B =\frac{9H^4}{2 \pi^2}$, however as already noted in the literature, in that case, the electric field becomes dominant leading to backreaction problems and invalidating the perturbative calculation presented and also invalidating the form of the current (\ref{eq:current}).
\paragraph*{Discussion of the parameters} We define the electric and magnetic spectral index $n_E$ and $n_B$ as $n_{E/B}\equiv \frac{\partial \log P_{E/B}}{\partial \log k}$. To avoid the backreaction problem and validate the scheme of pair production considered by taking an induced current of the form  (\ref{eq:current}), one needs to consider a scenario where the electric field does not dominate the dynamics: $n_E>0$ implying: $\sigma \in \left]-4,0\right]$. The strong coupling problem is also solved for these values. In that case $n_B \in \left]2,5\right]  $: the magnetic spectrum is blue, dominated by UV contributions and is more damped than its electric counter part. For those values, it has been shown however that the maximal magnetic field today at Mpc scales would be $10^{-32}~\mbox{G}$ \cite{Demozzi:2009fu}, which is lower than the observed lower bound \cite{Neronov:1900zz}. In IFF scenarios, the non-minimal kinetic term $I$ is sometimes parametrized as $I(\eta)\propto a^{n}$, in that case it would lead to the same equation of motion for the gauge field than the one considered in this work (\ref{eq:eom}) with $2n=\sigma$. This is an important result of this work as particle creation is expected as soon as the gauge field is present and our result is a clear backreaction results on how the gauge field dynamics would change in the presence of this particle: the spectral index of the magnetic field for instance would be shifted by $\frac{\sigma}{2}$.
We stress finally again that $\sigma$ cannot take too positive values as in that case the current $\mathcal{J}$ is proportional to $E^2$ only screening the establishment of the EM field.
\section{Conclusions} \label{sec:discussion}
\par Schwinger effect in de Sitter space has been widely investigated in the past years revealing new physical insights. For instance, renormalization schemes and their relation one to each other in curved spacetime is a developing field, see on that matter the recent works \cite{Ferreiro:2018qzr,Ferreiro:2018qdi}, the claim that the electromagnetic fields is of adiabatic order 1 in the presence of gravity requires that a lot of fundamental studies of the Schwinger effect \cite{Froeb2014,Kobayashi2014,Stahl:2015gaa,Stahl:2016geq,Hayashinaka:2016qqn,Hayashinaka:2016dnt,Bavarsad:2017oyv} have to be reviewed in the light of this. The processes involving quantum theory and gravity are also of crucial importance to write one day a complete theory of quantum gravity. Schwinger effect in dS also provides an interesting toy model to probe the inflationary response to the presence of an EM field. However in most of the studies on Schwinger effect, a constant electric field is assumed. An assumption that has its limitations \cite{Giovannini:2018qbq}. In this article, we explored the backreaction to the electromagnetic field. We showed that Schwinger effect impacts on the amplification of the EM quantum fluctuations. In \cite{Kobayashi2014}, it was already remarked that strong electric fields constrain a branch of the IFF scenarios in a similar manner than the backreaction problem. In this work, we completed this statement by considering negative conductivity (for weak electric fields) as a way to break the conformal invariance and to amplify EM quantum fluctuations. The presence of a negative conductivity is either motivated by a new implementation
of the adiabatic subtractions in the renormalization of the electric current presented in \cite{Hayashinaka:2018amz} or by the presence of non-linear terms generalizing the Einstein-Maxwell equation and a logarithmic running of the coupling constant $e$ \cite{Banyeres:2018aax}. We found models where both the strong coupling and the backreaction problem are avoided for $ \sigma \in \left]-4,0\right]$ however these models do not fulfill the lower bound of observed magnetic fields \cite{Neronov:1900zz}.
\par Those results are also relevant for primordial magnetogenesis model builder as the pair creation process depicted here matters as soon as EM fields are present during inflation. While in the case, of strong (backreacting) electric fields, one expects a dwindling of the EM field \cite{Kobayashi2014,Yokoyama:2015wws}, for weaker fields this effects could add up to any other way to break the conformal invariance. The global effect of Schwinger pair creation on the power spectrum of quantum fluctuations would be to have a redder spectral index, more concretely: $n_B \rightarrow n_B+\frac{\sigma}{2}$. Finally, we mention that it is desirable to include Schwinger effects for other primordial magnetogenesis setups such as phases transitions \cite{Vachaspati:1991nm,Sigl:1996dm}. Also during the reheating, many charged particles lead to a high conductivity and Schwinger effect could also take place changing the behavior of the electric field.

\end{document}